\documentclass{article}

\usepackage{arxiv}

\usepackage[utf8]{inputenc} 
\usepackage[T1]{fontenc}    
\usepackage{hyperref}       
\usepackage{url}            
\usepackage{booktabs}       
\usepackage{amsfonts}       
\usepackage{nicefrac}       
\usepackage{microtype}      
\usepackage{amsmath}
\title{Universal Area Law in Turbulence}

\author{
  Alexander Migdal \\
  Fresnel Research LLC
  } 
\begin{document}
\maketitle

\begin{abstract}
We re-visit the Area Law in Turbulence discovered many years ago \cite{M93} and verified recently in numerical experiments\cite{S19}. We derive this law in a simpler way, at the same time outlining the limits of its applicability. Using the PDF for velocity circulation as a functional of the loop in coordinate space, we obtain explicit formulas for vorticity correlations in presence of velocity circulation. These functions are related to the shape of the scaling function of the PDF as well as the shape of the minimal surface inside the loop. The background of velocity circulation does not eliminate turbulence but makes observable quantities in inertial range \textbf{calculable}. The scaling dimension of velocity circulation as a function of large area remains unknown. Numerical experiments \cite{S19} suggest transition for log-log derivative of circulation moments $\left<\Gamma^p\right>$ by the loop area from Kolmogorov index $\frac{2p}{3}$ at $p <4$ down to approximately $0.58 p$ for $4 \leq p \leq 10$ within available Reynolds numbers. We argue that Area Law applies to these moments only in the limit $p\rightarrow \infty$ when they are dominated by the tails of the PDF. So, these numerical experiments suggest that the scaling index in Area law is less then $\frac{2}{3}$.
\end{abstract}


\section{Introduction}
The strong turbulence looks like a hopelessly complex problem, with variety of multi-fractal models describing anomalous scaling laws for moments of velocity gradients \cite{Y07}. None of these models so far provides the microscopic picture nor allows computation of basic variables such as correlation functions of vorticity or velocity differences from the first principles. The root cause of these problems was the infrared divergency of the conventional approach based on Wyld diagram technique. The physical phenomenon behind this divergency is the growth of velocity correlation with distance. 

There is fluctuating velocity field in turbulent flow, generated by vortex structures which velocity moves these same structures resulting in complex long range interactions.
The statistics of these vortex cells was studied by conventional methods of local mechanics \cite{TSVS} which advanced our theoretical understanding but did not lead to any  observable predictions we would need from Turbulence theory.

A long time ago, in a Galaxy far, far away \cite{M93} we derived the loop equation which provided a geometric approach to Navier-Stokes equation, free of infrared divergencies. The Loop Equation had the form of Shroedinger Equation with nonlocal variables. Namely, the basic variable was a circulation around closed loop in coordinate space

\begin{equation}\label{Gamma}
    \Gamma = \oint_C \vec v d\vec r =\int_{S_C} \vec \omega d \vec \sigma
\end{equation}

The gradient part of velocity drops from the closed loop integral, like the gauge part of vector potential in electrodynamics. The incompressibility condition $\nabla \vec v =0$ is analogous to the gauge fixing condition in electrodynamics, but the circulation is gauge invariant just like in QED. 
In the same way the $\nabla p $ terms also drop from equations for circulation. 
The only relevant dynamical variable left is vorticity $\omega$, analogous to electromagnetic field. The role of the Planck's constant was played by viscosity, and the Reynolds number was the ratio of $\Gamma$ to viscosity $\nu$. Therefore, the turbulence corresponded to the WKB limit of the loop equation, opposite to the perturbative expansions in inverse powers of viscosity in the Wyld diagram method.

Let us dust off this old theory and present it in more comprehensible way. 

\section{Universal Area Law in Inertial Range}

It often happens that small observation outgrows the limitations of underlying assumptions and becomes a universal law. This is what seems to be happening with Area Law. When we tried to revive the original reasoning which led to Area Law we found out the way to derive it without assumptions of the WKB approximation in Loop Equations. Let us go step by step from definition to the Area Law.

The PDF  for  velocity circulation as a functional of the loop 

\begin{equation}
    P\left ( C,\Gamma\right) =\left < \delta\left(\Gamma - \oint_C \vec v d\vec r\right)\right>
\end{equation}

with brackets \begin{math}< > \end{math} corresponding to time average or average over random forces, was shown to satisfy certain functional equation (loop equation). 
Let us re-derive it here for the Euler equations
\begin{align*}
    \dot{v_i} &= -v_k \nabla_k v_i - \nabla_i p \\
    &= - v_k \omega_{k i} - \nabla_i \left(p + \frac{v_k^2}{2}\right);\\
    \omega_{k i} &= \nabla_k v_i - \nabla_i v_k
\end{align*}
 Integrating over the closed loop we find simpler formula, with gradient terns dropped 
 \begin{equation}
     \oint_C dr_i \dot{v_i} = -\oint_C dr_i v_k \omega_{k i}
 \end{equation}
Using incompressibility we can express velocity here as space integral of vorticity $\omega_{k i}$
\begin{equation}
    v_k(r) = \int d^3\rho\frac{ \rho_j }{4\pi|\vec \rho|^3}\omega_{k j}(r+\rho)
\end{equation}
Let us now compute the following mixed derivative of our PDF
\begin{align*}
  \frac{\partial}{\partial \Gamma} \frac{\partial}{\partial t}  P\left ( C,\Gamma\right) &=
   -\left < \delta''\left(\Gamma - \oint_C \vec v d\vec r\right)    \oint_C dr_i \dot{v_i} \right>\\
   &=\left < \delta''\left(\Gamma - \oint_C \vec v d\vec r\right)    \oint_C dr_i  \omega_{k i}(r) \int d^3\rho\frac{ \rho_j }{4\pi|\vec \rho|^3}\omega_{k j}(r+\rho)\right>
\end{align*}
Now, switching to vector from tensor vorticity:
\begin{equation}
    \omega_{a b}(r) = e_{a b c}\omega_c(r)
\end{equation}
and expressing vorticity as area derivative 
\begin{equation}
    \omega_k(r) = \frac{\delta}{\delta \sigma_k(r)} \int_{S_C} \vec \omega d \vec \sigma
\end{equation}
We finally arrive at the Euler limit of the loop equation:
\begin{equation}\label{LoopEq}
\frac{\partial}{\partial \Gamma} \frac{\partial}{\partial t}  P\left ( C,\Gamma\right) 
=\oint_C d r_i \int d^3\rho\frac{ \rho_j }{4\pi|\vec \rho|^3}\frac{\delta^2 P(C,\Gamma)}{\delta \sigma_k(r) \delta \sigma_l(r + \rho)}\left(\delta_{i j}\delta_{k l} - \delta_{i k}\delta_{j l}\right)
\end{equation}
The definition of area derivative $\frac{\delta}{\delta \sigma_i(r)} $ at the loop $C$ can be found in \cite{M93}, where some complicated limiting procedures were used. We are going to simplify that definition as follows.

Let us define area derivative using the difference between $U[C+\delta C]- U[C]$ where an infinitesimal loop $\delta C$ around the 3d point $r$ is added as an extra connected component of $C$. In other words, let us assume that the loop $C$ consists of an arbitrary number of connected components $C = \sum C_k$. We just add one more infinitesimal loop at some point away from all $C_k$. 
In virtue of the Stokes theorem, the difference comes from the circulation $\oint_{\delta C} \vec v d\vec r$ which reduces to vorticity at $r$
\begin{equation}
  P[C+\delta C]-P[C] =  d\sigma_i(r)    \left <\omega_i(r) \delta'\left(\Gamma - \oint_C \vec v d\vec r\right)\right>
\end{equation}
and in general, for the Stokes type functional, by definition:
\begin{equation}
    U[C+\delta C]-U[C] = d\sigma_i(r) \frac{\delta U[C]}{\delta \sigma_i(r)}
\end{equation}
The Stokes condition $\oint_{\delta S} d \sigma_i \omega_i =0$ for  any closed surface $\delta S$ translates into

\begin{equation}\label{Stokes}
   \oint_{\delta S} d \sigma_i \frac{\delta U[C]}{\delta \sigma_i(r)} =0
\end{equation}

The fixed point of the chain of the loop equations (\ref{LoopEq}) was shown to have solutions corresponding to two known distributions : Gibbs distribution and (trivial) global random rotation distribution. In addition, we found the third, nontrivial solution which is an arbitrary function of minimal area \begin{math} A_C \end{math} bounded by $C$.
\begin{equation}
    P(C,\Gamma) = F\left(A_C,\Gamma\right)
\end{equation}
The Minimal Area can be reduced to the Stokes functional by the following regularization
\begin{equation}
  A_C =\min_{S_C} \int_{S_C} d \sigma_{i}(r_1) \int_{S_C} d \sigma_{j}(r_2) \delta_{i j} \Delta(r_1-r_2)
\end{equation}
with 
\begin{equation}
\Delta(r) = \frac{1}{r_0^2} \exp\left(-\pi \frac{r^2}{r_0^2}\right); r_0 \rightarrow 0
\end{equation}
representing two dimensional delta function, and integration goes over minimized surface $S_C$. 

In general case the loop $C$ consist of $N$ closed pieces $C = \sum_{k=1}^N C_k$ and the surface $S_C$ must connect them all, so that it is topologically equivalent to a sphere with $N$ holes and no handles. In case some pieces are far away from others, the minimal surface would make thin tubes reaching from one closed loop $C_k$ to another via some central hub where all the tubes grow out of the sphere.
In our case we need only two extra little loops both close to the initial contour $C$ but for completeness we must assume there is an arbitrary number of closed pieces in $C$.

In real world this $r_0$ would be the viscous scale $\left(\frac{\nu^3}{\cal E}\right)^{\nicefrac{1}{4}}$.
This is a positive definite functional of the surface as one can easily verify using spectral representation:
\begin{equation}
    \int_{S_C} d \sigma_{i}(r_1) \int_{S_C} d \sigma_{j}(r_2) \delta_{i j} \Delta(r_1-r_2) \propto \int d^3k \exp\left(-\frac{k^2r_0^2}{4\pi}\right) \left| \int_{S_C} d \sigma_{i}(r)e^{i k r} \right|^2
\end{equation}
In the limit $r_0 \rightarrow 0$ this definition reduces to the ordinary area:
\begin{equation}
  A_C \rightarrow \min_{S_C} \int_{S_C} d^2 \xi \sqrt{g} 
\end{equation}

The Stokes condition (\ref{Stokes}) is satisfied in virtue of extremum condition, when the surface changes into $S'$, so that the linear variation reduces to the integral (\ref{Stokes}) with $\delta S = S'-S$ being the infinitesimal closed surface between $S'$ and $S$.

The area derivative of the Minimal Area in regularized form, then, as before, reduces to elimination of one integration
\begin{equation}
  \frac{\delta A_C}{\delta \sigma_i(r)} = 2 \int_{S_C} d \sigma_{i}(\rho) \Delta(r-\rho) \rightarrow 2 n_i(r)\exp\left(-\pi \frac{r_{\perp}^2}{r_0^2}\right)
\end{equation}
Where $n_i(r) $ is the local normal vector to the minimal surface at the nearest surface point $\Tilde{r}$ to the 3d point $r$, and $r_{\perp}$ is the component normal to the surface at $\Tilde{r}$. With this regularization area derivative is defined everywhere in space but it exponentially decreases away from the surface. Exactly at the surface it reduces to the unit normal vector.

Should we go to the limit $r_0 \rightarrow 0$ first we would have to consider the minimal surface connecting the original loop $C$ and infinitesimal loop $\delta C$. Such minimal surface would have an infinitely thin tube connecting the point $r$ to  $\Tilde{r}$ at the original minimal surface along its local normal $\vec n$. We are not going to investigate this complex problem here -- with the regularized area we have explicit formula, and we need this formula only at the boundary in leading log approximation (see below).

The second area derivative in (\ref{LoopEq}) reduces to derivatives with respect to $A_C$:
\begin{align*}
    \frac{\delta^2 F\left(A_C,\Gamma\right)}{\delta \sigma_k(r) \delta \sigma_l(r + \rho)} &=
    \frac{\delta}{\delta \sigma_k(r)} \frac{\delta A_C}{\delta \sigma_l(r + \rho)}\frac{\partial F}{\partial A_C}\\
    &=\frac{\delta^2 A_C}{\delta \sigma_k(r)\delta \sigma_l(r + \rho)} \frac{\partial F}{\partial A_C}
    +\frac{\delta A_C}{\delta \sigma_k(r)}\frac{\delta A_C}{\delta \sigma_l(r + \rho)} \frac{\partial^2 F}{\partial A_C^2} 
\end{align*}
The first term in the last expression is independent of the surface
\begin{equation}
    \frac{\delta^2 A_C}{\delta \sigma_k(r)\delta \sigma_l(r + \rho)}  = 2\delta_{k l} \Delta(\rho)
\end{equation}
so it does  not restrict integration $d^3\rho \rho_j/|\rho|^3$ and therefore vanishes due to the symmetry of $\Delta(\rho)$ with respect to rotations and reflection of $\rho$.
In the second term the constant factor $\frac{\partial^2 F}{\partial A_C^2} $ can be taken out of the integral so we get the following condition to be satisfied for the Area Law to be valid:
\begin{equation}
    \oint_C d r_i \int d^3\rho\frac{ \rho_j }{|\vec \rho|^3}\exp\left(-\pi \frac{\rho_{\perp}^2}{r_0^2}\right)n_k(r) n_l(r+\rho) \left(\delta_{i j}\delta_{k l} - \delta_{i k}\delta_{j l}\right)=0
\end{equation}
Let us stress the implied difference between these two terms. They both come from surface derivative of the Area, but the first variation, involving the normal vector $n_k$ is projected on a surface by the 2d delta function $\Delta(\rho)$.  Unlike that, the second derivative is a function of 3d vector $\rho$ and nothing else. So, integration goes over whole space and therefore vanishes by reflection symmetry. In a way it represents the expectation value of $\left<v_i(r) \omega_{ij}(r)\right>$ which vanishes due  to space symmetry. I hope mathematicians will find the better  way to formalize these vague arguments.

Note that we \textbf{did not} make any assumption of decay of PDF, unlike the old paper \cite{M93}. However, there are implied assumptions about large loop and large velocity circulation $\Gamma$.

The integral logarithmically diverges at small $\rho$, as we shall see soon, so we can replace $n_k(r+\rho)\rightarrow n_k(r)$ and take out of the integral over $d^3 \rho$.
We find then:
\begin{equation}\label{NNintegral}
    \oint_C d r_i \int d^3\rho\exp\left(-\pi \frac{\rho_{\perp}^2}{r_0^2}\right)
    \frac{ \rho_i - n_i(r) (\vec n(r)\vec \rho)}{|\vec \rho|^3}
\end{equation}
The integration over orthogonal component $\rho_{\perp}$effectively projects $\rho$ on the surface. Dropping irrelevant factor $r_0$ resulting from this integration we get surface integral (with $t_i(\theta) = C_i'(\theta)$ being local tangent vector to the loop)
\begin{equation}
    \int_0^{2\pi} d \theta  \int d^2\rho\frac{ \rho_i t_i(\theta) }{|\vec \rho|^3}
\end{equation}
This integral goes over semi-plane corresponding to inside of the surface. In a frame where $t_i(\theta)$ is directed along $x$ and the inside direction to the surface  is  $y$, the normal vector $n_i$ is directed along $z$. Now the integral of $\rho_x$ vanishing in the leading approximation of the semi-plane.
Thus, the leading term vanishes.

Note that without area derivative vector $n_i$ being directed along the normal to the minimal surface the logarithmically divergent term would not cancel:
\begin{equation}\label{logdiv}
    \int d^2\rho\frac{ \rho_i n_i }{|\vec \rho|^3} =
    \int_{0}^{\infty} dy\int_{-\infty}^{\infty} dx \frac{ x n_x + y n_y }{\left(x^2 + y^2\right)^{\nicefrac{3}{2}}} \propto \int_{0}^{\infty} \frac{dy n_y}{y} \propto -n_y \log(0)
\end{equation}

We computed the integral over $\rho$ with logarithmic accuracy at $\rho \rightarrow 0$, which means that we only needed area derivative directly at the loop. 
We neglected next terms which are finite, because the leading logarithmic terms cancelled. These finite terms involve global surface integrals and would need more sophisticated method to compute. The balance of these terms will determine the next corrections to the Area law.
In the next approximation we would have to add higher corrections both to the loop in vicinity of integration point and to the areas $A_C$ like
\begin{equation}\label{ExtraTerms}
   \delta A_C = \int_{S_C} d \sigma_{i}(r_1) \int_{S_C} d \sigma_{j}(r_2) T_{i j k l}\left(r_1^k- r_2^k\right)\left(r_1^l- r_2^l\right)\Delta(r_1-r_2) + \dots
\end{equation}
where integration goes over the minimal surface $S_C$ and $T_{i j k l}$ is some symmetric tensor made of Kronecker deltas. These terms in area derivative vanish at the boundary to preserve the condition:
\begin{equation}
    \frac{\delta A_C}{\delta \sigma_i} \rightarrow 2 n_i
\end{equation}
at the boundary of the surface, which we used to cancel the leading logarithmic divergency in the loop equation(\ref{LoopEq}).

In this paper we are only studying the Area Law and its consequences.

As we argued in original paper \cite{M93} we expect scale invariant solutions, depending of $\gamma =\Gamma A_C^{-\alpha}$ in our scale invariant equations, with some critical index, yet to be determined.
Let us stress again, that the Kolmogorov value of the scaling index $\alpha_{K} = \frac{2}{3}$ does \textbf{not} follow from the loop equations, this is an additional assumption, based on dimensional counting and Kolmogorov anomaly \cite{M93} for the third moment of velocity. As it was stressed in that paper, the Kolmogorov anomaly poses no restrictions on the vorticity correlations, and cannot therefore be used to determine our scaling index. 

The normalized PDF reads:
\begin{equation}\label{ScalingAreaLaw}
    P(C,\Gamma) = A_C^{-\alpha} \Pi\left(\Gamma A_C^{-\alpha}\right) 
\end{equation}
\begin{equation}
    \int_{-\infty}^{\infty}d\gamma\Pi\left(\gamma\right) =1 \label{Norm}
\end{equation}

This Universal Area Law  was confirmed in numerical experiments \cite{S19} with Reynolds up to $10^4$ with very high accuracy over whole inertial range of circulations and areas with PDF from 1 down to $10^{-8}$. 
To be more precise, in these experiments universality applies to dimensionless quantities, such as PDF in terms of $\Gamma \left<\Gamma^2\right>^{-\nicefrac{1}{2}}$. As for the second moment $\left<\Gamma^2\right>$ as well as all other fixed moments, these are not the universal functions of the area. 

Simple example\footnote{I am grateful to Polyakov and Yakhot for pointing this out to me.}  of the second moment $\left < \Gamma^2 \right> $ as computed with K41 scaling law for velocity correlation, manifestly depends upon aspect ratio of rectangular loop. 
According to \cite{S19} this dependence disappears in scale invariant ratios of moments $\left<\Gamma^n\right> \left<\Gamma^2\right>^{-\nicefrac{n}{2}}$.
In our scaling variable $\gamma$ this means that some non-universal scaling factor multiplies the power of area.

Theoretically, any finite moment is determined by a multiple vector integral over loop. This is an obviously different functional of the loop then the minimal area which is a non-polynomial functional of the loop.
In our old paper \cite{M93} we never claimed that area law applied to the circulation moments, but rather only to the tails of the PDF at large $\Gamma$.

At the closer look at the data presented in \cite{S19} we notice that there is indeed a verification of the independence of the PDF upon the aspect ratio of the rectangular loop in wide range of these ratios. However the PDF in log scale is totally dominated by its tail so one cannot conclude from their pictures that  region of small $\Gamma$ is as universal as the tails. The moments data, though were not studied as functions of the aspect ratio, only their dependence of the Area was presented.

So, the careful conclusion from \cite{S19} is that area law holds for PDF tails in surprisingly wide range of the scaling variable $\gamma = \Gamma A_C^{-\alpha}$ and that it asymptotically decays as $ \exp\left(-b_\pm |\gamma|\right)$ . As for the effective scaling index 
\begin{equation}
    \alpha(p) = \frac{d \log \left<\Gamma^p\right>}{d \log A_C^p}
\end{equation}
in these experiments it varies from $\frac{2}{3}$ for small $p$ to $\alpha(10) \approx 0.58$ . Apparently, even larger Reynolds numbers would be needed to find true asymptotic value of $\alpha$. We discuss this important question in the Discussion section of this paper.

The main conclusion is that \textbf{Area Law really works} in isotropic turbulence at large Reynolds numbers.

This is quite encouraging! Now we can take these old predictions seriously and start investigating the applicability and consequences of Scaling Area Law. If you ask me what was I waiting for all these years -- I was waiting for any sign of interest to this theory from my colleagues and any indications that these laws are observable at least in numerical experiments. In a retrospect I am sorry for my lack of persistence and I feel sad that scientific community did not show more curiosity 26 years ago. 

\section{Vorticity Distribution Along the Loop}

Here comes the first new prediction. 
Let us use the following identity:

\begin{equation}
    \left < \left(\vec \omega \frac{\partial}{\partial \Gamma}  + \frac{\delta}{\delta \vec \sigma} \right)\delta\left(\Gamma - \int_{S_C} \vec \omega d\vec \sigma \right) \right> =0
\end{equation}

In the first term we can take \begin{math} \frac{\partial}{\partial \Gamma} \end{math} out of the averaging and in the second term we can take out \begin{math} \frac{\delta}{\delta \vec \sigma} \end{math} 

We also use the above formula for area derivative at the loop:

\begin{equation}
    \frac{\delta A_C}{\delta \vec \sigma} =2 \vec n
\end{equation}

with \begin{math} \vec n \end{math} being local normal vector to the minimal surface at the boundary \begin{math} C \end{math}.
We shall use the following notation for statistical average conditional of fixed velocity circulation:

\begin{equation}
    \left <  X \right >_{\Gamma,C} == \left <X \delta\left(\Gamma - \int_{S_C} \vec \omega d\vec \sigma \right) \right> 
\end{equation}

In particular, 

\begin{equation}
    \left <  1 \right >_{\Gamma,C} == \left < \delta\left(\Gamma - \int_{S_C} \vec \omega d\vec \sigma  \right) \right> 
    = P\left(A_C,\Gamma\right)
\end{equation}

As a result we obtain following relation for the expectation value of vorticity at the loop:

\begin{equation}\label{FirstMean}
    \frac{\partial}{\partial \Gamma} \left < \vec \omega \right >_{\Gamma,C} = - 2\vec n \frac{\partial}{\partial A_C} P\left(A_C,\Gamma\right)
\end{equation}

Finally, after integrating over $\Gamma$ we obtain expected value of vorticity:
 
 \begin{equation} \label{FirstOmegaMoment}
     \left < \vec \omega \right >_{\Gamma,C} =2\vec n \frac{\partial}{\partial A_C}  \int_\Gamma^{\pm\infty}d \Gamma' P\left(A_C,\Gamma'\right)
 \end{equation}

The higher correlation functions of vorticity can be studied by the same method, by using recurrent relations:
 
\begin{equation}\label{RecurrentEquations}
 \frac{\partial}{\partial \Gamma}   \left < \vec \omega_1...\vec \omega_k \vec \omega_{k+1} \right>_{\Gamma,C}  +2 \vec n_{k+1} \frac{\partial}{\partial A_C}\left < \vec\omega_1...\vec\omega_k\right>_{\Gamma,C} =0
\end{equation}

with \begin{math} \omega_k, n_k \end{math} denoting vorticity and normal vectors at different points on the loop. 
The coordinate dependence factors out (assuming general index $\alpha$ in scaling variable):

\begin{equation}\label{MultiNormal}
    \left < \vec \omega_1\dots \vec\omega_k\right >_{\Gamma,C}  = \vec n_1 \dots \vec n_k A_C^{-\alpha - k(1-\alpha)}\Omega_k\left(\Gamma A_C^{-\alpha}\right)
\end{equation}

These equations reduce to the set of recurrent for multiple correlations  $\Omega_k\left(\gamma\right)$
of the form:
\begin{equation}
    \Omega_{k+1}\left(\gamma\right)= 2 \alpha \gamma\Omega_{k}\left(\gamma\right)  -2(1-\alpha)k\int_\gamma^{\pm \infty}\Omega_{k}(y) \,d y
\end{equation}
\begin{equation}\label{Omega0}
    \Omega_0(\gamma) = \Pi(\gamma)
\end{equation}
With help of $Mathematica^{TM}$, we find the following:

\begin{equation}\label{Omega1}
    \Omega_1(\gamma) = 2 \alpha\gamma \Pi(\gamma)
\end{equation}
\begin{equation}\label{Omega2}
    \Omega_2(\gamma) = \left(2 \alpha\gamma\right)^2\Pi(\gamma) -4 \alpha(1-\alpha)\int_{ \gamma}^{\pm\infty} \Pi(y)y \,d y 
\end{equation}
\begin{equation}\label{Omega3}
    \Omega_3(\gamma) =(2\alpha \gamma)^3
   +8  \alpha (1-\alpha ) \int_{\gamma }^{\pm \infty } (2 (1-2
   \alpha ) y-(2-\alpha ) \gamma )  \Pi (y) y\, d y
\end{equation}

It is understood that integrals $\int_{\gamma}^{\pm\infty}$ go to $+\infty$ in case $\gamma >0$ and $ -\infty$ otherwise. Given experimental data \cite{S19} for $\Pi(\gamma)$ these integrals can be computed with high accuracy, as there are no cancellations and exponential convergence. We have to rely on experiments until the theory will advance to provide us exact form of $\Pi(\gamma)$. 

Note that the vorticity correlations are strongly correlated (lack of correlation would correspond to $\Omega_2\Omega_0 \rightarrow \Omega_1^2$). Surely they are not Gaussian either -- that would imply vanishing odd correlations and algebraically related fourth and second correlations which is not the case.

The first terms in (\ref{Omega2}),(\ref{Omega3}) represent disconnected part of expectation values, the rest representing correlations. Note that asymptotically, at large $\gamma$ these correlation terms decay faster than the leading term. This is because the $\Pi(\gamma)$ decays as $\exp\left(-b |\gamma|\right)$ so that integrals are dominated by vicinity $y-\gamma = O(1)$ which leads to $ \gamma^{-1}$ estimate of the correlation terms compared to the leading first term.

Note that these are \textbf{not} the moments of powers of vorticity at the same point -- we do not know how to compute those nor do we know if those moments obey Kolmogorov scaling law. We are computing expectation values of products of vorticity at different points at the loop, separated by more than viscous scale distance.

In particular, at zero circulation
\begin{equation}\label{Omega2p}
    \Omega_2(0^+) = -4 \alpha(1-\alpha)\int_{ 0}^{\infty} d \gamma \gamma\Pi(\gamma)
\end{equation}
\begin{equation}\label{Omega2m}
    \Omega_2(0^-) = -4 \alpha(1-\alpha)\int_{-\infty}^{0} d \gamma |\gamma|\Pi(\gamma)
\end{equation}
Are these integrals equal? Numerical experiments can tell. This discrepancy is one more indication that our area law does not apply to the region of small $\Gamma$ which is essential here.

Another interesting subject is two loop PDF, where the contour $C$ consists of two separate loops $C_1, C_2$ and we separately fix the circulations $\Gamma_1, \Gamma_2$ on each of them. In general the double loop PDF
\begin{equation}
     P_2\left ( C_1,\Gamma_1;  C_2,\Gamma_2\right) =\left < \delta\left(\Gamma_1 - \oint_{C_1} \vec v d\vec r\right)\delta\left(\Gamma_2 - \oint_{C_2} \vec v d\vec r\right)\right>
\end{equation}
can be represented as sum of connected and disconnected parts:
\begin{equation}
    P_2\left ( C_1,\Gamma_1;  C_2,\Gamma_2\right) = P\left ( C_1,\Gamma_1\right) P\left( C_2,\Gamma_2\right) + \Phi\left ( C_1,\Gamma_1;  C_2,\Gamma_2\right)
\end{equation}
\begin{equation}
    \int d \Gamma_1 \Phi\left ( C_1,\Gamma_1;  C_2,\Gamma_2\right) = \int d \Gamma_2 \Phi\left ( C_1,\Gamma_1;  C_2,\Gamma_2\right) =0
\end{equation}
where $\Phi$ represents the loop-loop correlation function. 

Note that both terms would satisfy the loop equations (\ref{LoopEq}), but they correspond to different topologies of the minimal surface: in the first term the surface represents two disconnected discs bounded by $C_1$ and $C_2$ and in the second one the surface connects these two loops.

We expect that this second term depends on the area $A(C_1,C_2)$ of the minimal surface connecting these loops, like a tube. The scaling law would look as follows
\begin{equation}
    \Phi\left ( C_1,\Gamma_1;  C_2,\Gamma_2\right) = A(C_1,C_2)^{-2  \alpha} \phi\left(\Gamma_1 A(C_1,C_2)^{-\alpha} ,\Gamma_2 A(C_1,C_2)^{-\alpha}\right)
\end{equation}
In particular, at large separations between the loops the minimal area $A(C_1,C_2)$ would grow linearly with distance $r_{1 2}$ as the minimal surface shrinks to a thin tube. In that limit we expect the scaling function $\Phi$ to reach finite limit as zero arguments so we would get a usual scaling law:
\begin{equation}
    \Phi\left ( C_1,\Gamma_1;  C_2,\Gamma_2\right) \rightarrow A(C_1,C_2)^{-2\alpha } \phi\left(0,0\right) \propto r_{1 2}^{-2\alpha}
\end{equation}

\section{Discussion}
The above theory goes against many popular beliefs as well as established facts regarding anomalous scaling of the moments of velocity differences and velocity gradients \cite{Y07}.

Here is my response to questions from one of the best experts in that field who asked me how my simple scaling laws and nonsingilar spatial dependence of vorticity correlations can be compatible with multi-fractal structure of velocity gradients.

There is a lot of complexity and anomalous scaling in velocity differences and velocity gradients. 
However, vorticity = curl v is a very special case of velocity gradient, and this is the only one I address. 
Plus, as I already mentioned, I am considering strong background of velocity circulation around specific loop, which violates space symmetries you had in your case.

This is why mean vorticity at the loop point is not zero, as it would have been in ordinary unconstrained turbulent flow, it is rather directed at the normal to the minimal surface at this point at the loop.
It also goes to zero as $A_C^{\alpha-1}$ according to our scaling law. 

It is worth mentioning that all my results assume velocity circulation $\Gamma$ large compared with viscosity $\nu$ and the minimal area inside of the loop $C$ large compared to viscous scale
$ r_0 = \left(\frac{\nu^3}{\cal E}\right)^{\nicefrac{1}{4}}$.

There are strong indications \cite{Y18}, that rather then a single Kolmogorov spatial scale $r_0$ there are many scales corresponding to changes from one multi-fractal regime to another, in other words, strong intermittency.
The circulation PDF measurements \cite{S19} showed much less intermittency in our case. There is a relatively slow dependence of log-log derivative of circulation moments with respect to the minimal area. 

We have to stress that lower moments (nor any finite moments for that matter) are not expected to be universal functions of the minimal area even after normalizing by the root of the second moment. Our derivation assumed large circulation, where as any finite moments are influenced by the region of small circulation where we cannot expect area law. 

In the next paper we argue that $\alpha = \frac{1}{2}$ based on certain self consistency relations and compare this prediction with experimental data \cite{S19}.

So, I found a quiet harbor in a stormy sea of turbulence, namely large velocity circulation PDF, which seems to be calculable and seems to agree with experiments (even better than I initially expected).
I am not claiming the the whole minimal surface has any direct physical meaning -- like the dominant configuration of vorticity along this surface with viscous thickness. 

Such hypothesis would explain my formulas, but it is not necessary to obtain them. 
In the same way as Gibbs distribution does not tell us anything about actual kinetics leading to such distribution, the Area law does not mean the dominant vorticity configurations in a turbulent flow follow minimal surface.

In case of Gibbs, actual dynamics could take a lot of time to uniformly cover energy surface in phase space. There are examples of fake dynamics such as Langevin equation  $\dot r = - \nabla U + f$ with Gaussian random force $f$ leading to Boltzmann distribution.

There may be many kinds of evolutions leading to the same PDF satisfying Area law.
In the same way, as complexity of classical mechanics of N bodies does not mean random motion along the energy surface, the turbulent dynamics may uniformly cover minimal surface without being concentrated along this surface.

This minimal surface may be only a solution of equation for equilibrium PDF, but not a solution to actual Euler dynamics in space-time. We have completely different physical picture in equilibrium PDF and actual turbulent dynamics. In particular, the Stokes condition (\ref{Stokes}) which comes as an identity in fluid dynamics becomes a dynamic relation here, namely extremum condition of the minimal surface.  Duality between these two pictures looked like science fiction 26 years ago, and would probably still look like that were it not for remarkable numerical experiments \cite{S19} vindicating that moonshot.

 \section{Conclusions}
 
 We have found rich structure in statistical theory of velocity circulation in strong isotropic turbulence. Within the same WKB assumptions which led us to Scaling Area Law (\ref{ScalingAreaLaw}) we now computed the vorticity expectation values along the loop (\ref{Omega1}). We also found integral expressions (\ref{Omega2}) for second correlation function. Recurrent equations (\ref{RecurrentEquations}) express all higher correlations in terms of integrals involving (\ref{Omega0}).
 
 There are two kinds of relations here, namely Universal Area Law which follows from evolution equation for circulation PDF and on top of that, an assumption about  scaling for PDF. We did not provide any dynamical arguments in favor of scaling except for scale invariance of the loop equations.
 
 Maybe it is worth stressing that recurrent equations (\ref{RecurrentEquations}) are exact asymptotic formulas for vorticity correlations in presence of large velocity circulation in strong isotropic turbulence. The solutions  for first three correlations were derived under further assumption of scaling law. 
 Neither the old paper \cite{M93} nor this one makes any claims about lower moments. Area Law was derived only for the tails of PDF, which are responsible for the higher moments. 
 
The dimensionles combinations 
\begin{equation}\label{UniversalComb}
    \xi(p) =\frac{1}{p}\log\left<|\Gamma|^p\right> - \frac{1}{p-1}\log\left<|\Gamma|^{p-1}\right> 
\end{equation}
must become independent of aspect ratio at large $p$. At $p=3$, for example, we expect some dependence of the aspect ratio, though we cannot compute it as the three-point velocity correlation function is unknown.
 
 The minimal surface is the most important immediate source of predictions for the vorticity correlations. Minimal surface is  a classical Plateau problem \cite{MinS} which we revisited in Appendix in \cite{M93}. It can be computed for any shape of the loop either analytically or numerically.  The same applies therefore to vorticity correlations  as functions of a point on a loop: they are all proportional to product of corresponding normal vectors (\ref{MultiNormal}).
 
 Much more challenging problem is to explain the whole profile of  PDF function $\Pi(\gamma)$ measured in \cite{S19}. 
 
 \section{Suggested Numerical Experiments}
 It would be very interesting to verify these predictions in numerical experiments. 
 To be more specific, I would suggest the following numerical experiments and computations.
 
\begin{itemize}
\item Measure dependence of the dimensionless variables (\ref{UniversalComb}) of the aspect ratio of rectangular loop.
\item Measure the scaling exponents of higher moments for larger order $p >10$.
\item Compare the numerical results for the second moment $\left<\Gamma^2\right>$ for a rectangular loop with result of numerical integration of K41 velocity correlation over the loop $ \oint_C d \vec r_1 \oint_C d \vec r_2 |\vec r_1 -\vec r_2|^{\frac{2}{3}}$.
\item Measure vorticity mean value (\ref{Omega1}) for the flat square loop as function of scaled circulation $\gamma=\Gamma A_C^{-\alpha}$.
\item Measure second correlation (\ref{Omega2}) for two points on square loop as the function of $\gamma$.
\item In particular compare limits (\ref{Omega2p}), (\ref{Omega2m}).
\item Take two perpendicular unit squares size $a$: one in $x y$ plane, 
another in $y z$ plane, touching along $y$ side, like a soccer gates. Each square has calculable circulation and total circulation is just a sum of the two (the wires on common side cancel each other).
However, the Area Law predicts quite a nontrivial curved minimal surface $S_C$ which is well studied in the literature \cite{MinS}. So, measure the first correlation as a function of the point on a loop. It will be directed normal to this minimal surface, which is very far from the original planes where the squares belong. Check this vorticity vector rotation along the loop.
\item Study the transition between area law and power law as the minimal surface between two squares in $ x y $ plane separated by a distance $r_{1 2}$ in $z$ direction shrinks to a thin tube.
\item It would be nice to compare our formula (\ref{Omega3}) for the third correlation with numerical experiments at least for a simple square.
\end{itemize}
\section{SOS}
 The Loop Equation in principle can be iterated to find next corrections to Area Law. The dynamical variables here are loop functionals, so some string theory techniques may apply. Where are you, bored string theorists, looking for a beautiful applications of your mathematical wisdom? Come and help us! The  Statistical Circulation Theory is just being reborn, best discoveries are still waiting for us.
 
 \section{Acknowledgements}
 I am indebted to Katepalli R. Sreenivasan and Victor Yakhot for stimulating discussions. My renewed interest to this subject came about because of remarkable numerical experiment Sreeni made with his collaborators \cite{S19}. The seminar discussion I had afterwards few days ago in his Brooklyn office inspired me to generalize old asymptotic area law to the Universal Area Law as I see it now. Long and deep discussion with Sasha Polyakov helped me understand the meaning of area derivative away from the boundary of the surface.
I am also grateful to Mitch Sonies and my team in Fresnel Research LLC for their moral support of my scientific work.
\bibliographystyle{unsrt}  


\end{document}